\documentclass[aps,prl,twocolumn,superscriptaddress,longbibliography,numerical]{revtex4-1}

\usepackage{amsmath}
\usepackage{epsfig}
\usepackage{graphicx}
\usepackage{bm}
\usepackage{amssymb}
\usepackage{slashed}
\usepackage{hyperref}
\hypersetup{
     colorlinks   = true,
     citecolor    = blue,
     urlcolor = blue,
     linkcolor = blue
}
\newcommand{\bsigma}{{\boldsymbol \sigma}}

\begin{document}

\title{Spin and valley waves in Dirac semimetals with population imbalance}

\author{A.~A.~Zyuzin}
\affiliation{Department of Applied Physics, Aalto University, P.~O.~Box 15100, FI-00076 AALTO, Finland}
\affiliation{Ioffe Physical--Technical Institute,~194021 St.~Petersburg, Russia}

\author{A.~Yu.~Zyuzin}
\affiliation{Ioffe Physical--Technical Institute,~194021 St.~Petersburg, Russia}

\begin{abstract}
We find an intervalley wave collective mode in two- and three-dimensional Dirac semimetals in the presence of a valley population imbalance.  
The dispersion relation of this mode is gapless, proportional to the square of the wave vector at small frequencies, and inversely proportional to the electron-electron exchange interaction energy.
The valley wave serves as an energy gain source for the external field, that generates the intervalley transitions. The spin wave analog is discussed for the case of a semimetal with nonequilibrium spin orientation.
\end{abstract}
\maketitle

The theory of spin waves in metals and semiconductors with non-equilibrium spin orientation was pioneered by Aronov in Ref. \cite{Aronov} and later developed, for example, in Refs. \cite{Bashkin, Rediscovery, Zyuzin_Zyuzin, Bashkin_review}. 
Utilizing the methods of optical spin orientation or electric current spin injection, one can tune a paramagnetic material in a situation with an unequal population of spin states. 
In this case, due to an electron-electron exchange interaction, a gapless spin wave mode with a quadratic dispersion law can propagate in the system. The collective spin excitations were experimentally detected
in gaseous spin-polarized hydrogen, polarized $^3\textrm{He}$, and in mixtures of $^3\textrm{He}$ - $^4\textrm{He}$ (see Ref. \cite{Bashkin_review} for a review) as well as in a magnetically trapped ultracold atomic gas  \cite{Experiment_spinwaves2}.

In analogy to spin, an electronic valley degree of freedom in multivalley two-dimensional (2D) hexagonal materials, such as graphene and transition-metal dichalcogenides or three-dimensional (3D) Weyl-Dirac semimetals, may give rise to collective valley waves. Moreover, the valley pseudospin degree of freedom can be utilized in a thin-film topological insulator hosting Dirac surface states.

It was proposed that the valley imbalance in graphene can be induced, for example,
upon injection of electric current through a ballistic point contact with zigzag edges \cite{Been}, by optical radiation due to warping of the electron energy spectrum \cite{Optical_separation_graphen}, or by time-dependent mechanical deformations \cite{AC_deformations}. To the best of our knowledge, indications of valley currents in graphene are missing.
Although, in systems with broken spatial inversion symmetry, such as monolayer transition-metal dichalcogenides and a biased graphene bilayer, valley-dependent optical selection rules allow the formation of a valley imbalance.
For a review, see Ref. \cite{Valleytronics}. In 3D Weyl-Dirac semimetals the valley imbalance can be induced via a chiral anomaly, which implies that the application of collinear electric and magnetic fields pumps electrons between nodes \cite{Nielsen}.

Here, we find the valley and spin wave collective modes in nonequilibrium doped Dirac semimetals under the condition of valley and spin population imbalances, respectively. 
The mass of the waves is proportional to the electron-electron exchange interaction energy. The quadratic wave vector dependence in the dispersion relation is different from the square-root and linear dependencies
of the intra and inter valley plasmon modes in doped graphene \cite{RevModPhys_graphene_2, RevModPhys_spectroscopy, Intervalley_plasmon}, as well as from the gapped mode in 3D Weyl-Dirac semimetals \cite{Kharzeev}. 

\emph{Collective modes in a 2D semimetal with population imbalance}.
We proceed with the analytical derivation of the dispersion relation for the valley and spin waves in doped graphene in a situation with valley and spin imbalances (these calculations can be equally applied to the surface states of a thin-film topological insulator). Let us start with the valley waves.

The linearized low-energy model of two spin-degenerate inequivalent Dirac points in graphene is described by the Hamiltonian 
\begin{eqnarray}\label{Graphene_Hamiltonian}
H_{\eta, ab}(\mathbf{p}) = v(\eta \sigma^x_{ab} p_x + \sigma^y_{ab} p_y) - \eta \frac{\lambda n}{2} \sigma^0_{ab},
\end{eqnarray}
where $v$ is the Fermi velocity, $\sigma^{0}$ is the unit matrix and $\sigma^{x,y}$ are the Pauli matrices acting on sublattice space with matrix indices $(a,b)$ suppressed here for brevity, $\eta=\pm $ denotes the valley index, and momentum $\mathbf{p}$ is counted from the position of the corresponding Dirac point. We use $k_B= \hbar = 1$ units henceforth.

The third term describes the nonsymmetric part of the electron-electron exchange interaction energy between two valleys, where the unimportant identical shift is already included in the chemical potential.
We consider the simplest approximation for the Fourier component of the screened potential of the exchange interaction $\lambda$ by taking it to be momentum independent. 

\begin{figure}[t]
\begin{tabular}{c}
\centering
\includegraphics[width=8cm]{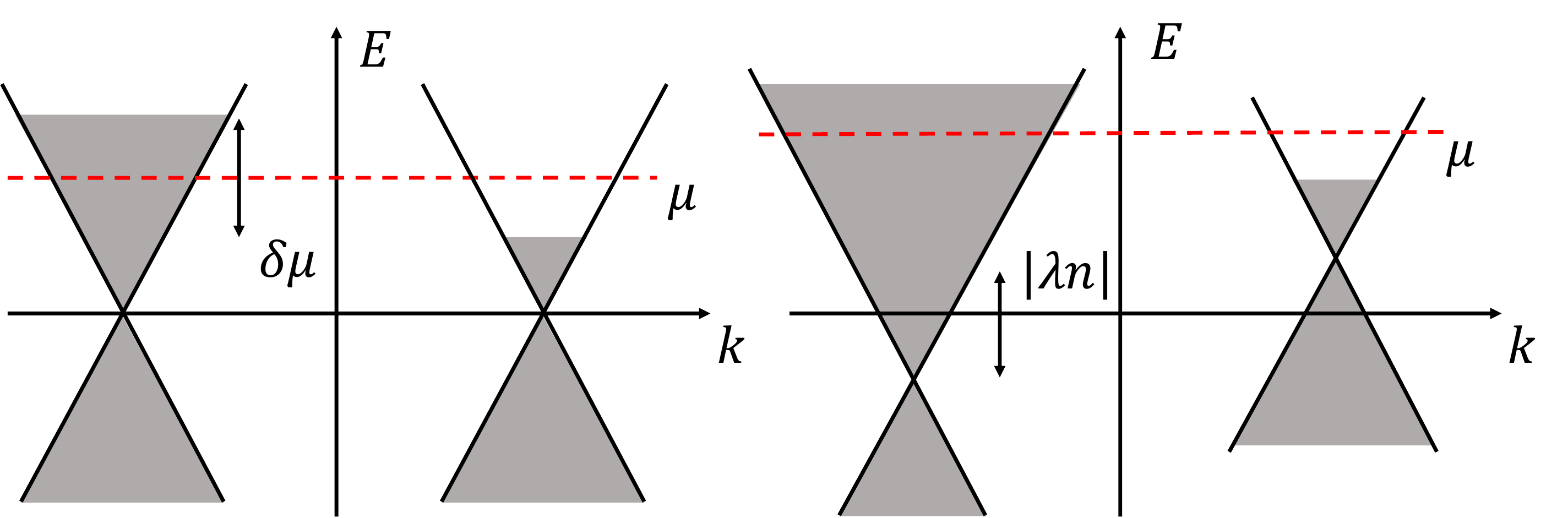}
 \end{tabular}
\caption{Schematic illustration of the valley imbalance in a doped Dirac semimetal. Left: In the absence of an electron-electron interaction, the population imbalance is described by the relative shift of the chemical potential $\mu \pm \delta \mu/2$ in two Dirac valleys. 
Right: Turning on the intervalley exchange interaction in the presence of imbalance, two Dirac points are shifted with respect to each other by the amount of exchange energy $|\lambda n|$, where $\lambda$ is the interaction constant and $n$ is the particle density difference between two valleys induced by the imbalance. }
\label{fig1} 
\end{figure}

The exchange energy in Eq. (\ref{Graphene_Hamiltonian}) is determined by the relative valley pumping and shall be found self-consistently. The particle density difference between the valleys per one sublattice is given by
\begin{equation}
n = \frac{g}{2}\sum_{s=\pm}\int \frac{d^2 p}{(2\pi)^2} \left\{ f_{+}[E_{s,+}(\mathbf{p})] - f_{-}[E_{s,-}(\mathbf{p})] \right\},
\end{equation}
where
$
f_{\eta}(E_{s,\eta}) = [1+e^{(E_{s,\eta}-\mu_\eta)/T}]^{-1}
$
is the sublattice-independent Fermi-Dirac distribution function in valley $\eta$ with $T$ as temperature, 
\begin{equation}
E_{s, \eta}(\mathbf{p}) = s v |\mathbf{p}| - \eta \frac{\lambda n}{2} 
\end{equation}
is the spectrum of particles, $s=\pm$ labels the conduction and valence bands of the Dirac cone, and $g=2$ is due to the summation over two spins. 
The imbalance is described by the relative shift of the chemical potential in two valleys as 
\begin{equation}
\mu_{\eta} = \mu + \eta \frac{\delta\mu}{2}. 
\end{equation}
In the present model, in the absence of an interaction, the $\delta\mu$ does not change the spectral properties of the Dirac fermions. Turning on the interaction, the exchange energy $\lambda n$ becomes finite and splits the Dirac cone in energy, as shown in Fig. \ref{fig1}.

Indeed, let us consider the case of a doped semimetal with the chemical potential set in the conducting band $\mu>0$ and assume that conditions $\mu \gg |\delta\mu| $ and $\mu\gg |\lambda n|$ are satisfied. Hence, the renormalized particle density imbalance is given by
 \begin{equation}
 n = \frac{\mu}{2\pi v^2} \frac{\delta\mu}{1- \frac{\lambda \mu}{2\pi v^2} }.
 \end{equation}
The relative energy shift of Dirac points, given by $|\lambda n|$ in Eq. (\ref{Graphene_Hamiltonian}), vanishes with the decrease of both the valley imbalance and the strength of the electron-electron interaction. 
Also note that the sign of $n$ coincides with the sign of $\delta\mu$ as well as that the exchange energy is smaller than the relative shift of the chemical potential $|\lambda n| < |\delta\mu|$. The spectrum and the population imbalance are schematically shown in Fig. \ref{fig1}.

Consider now the system in the presence of the external field $\propto e^{i\mathbf{q}'\cdot \mathbf{r} -i\omega t}$,  that causes transitions between the states in different valleys, i.e., generates the off-diagonal matrix elements in the valley pseudospin space. Here, the length of the wave vector $\mathbf{q} = \mathbf{q}' - \mathbf{K}$ is much smaller than the length of the one that connects the two valleys $\mathbf{K}$, 
such that $| \mathbf{q}' - \mathbf{K}| \ll \mu/v \ll K$. The response functions shall be studied on the time-scales shorter than the typical inelastic intravalley and elastic intervalley relaxation times. 

The inter-valley processes are described by the polarization operator, in which two Green functions are from the opposite valleys.
The retarded and advanced Green functions are given by 
\begin{eqnarray}\label{RA}
G^{R,A}_{\eta, ab}(\mathbf{p},\omega) =\frac{1}{2}\sum_{s=\pm}\frac{\sigma^0_{ab} + s (\eta \sigma^x_{ab}\hat{p}_x + \sigma^y_{ab}\hat{p}_y)}{\omega -E_{s, \eta}(\mathbf{p})  \pm i\delta},
\end{eqnarray}
where $\hat{\mathbf{p}} = \mathbf{p}/| \mathbf{p}|$ is the unit vector in the direction of the momentum. The Keldysh Green function reads 
\begin{eqnarray}\label{K}
G^{K}_{\eta, ab}(\mathbf{p},\omega) = [1-2 f_{\eta}(\omega)]  [G^{R}_{\eta, ab}(\mathbf{p},\omega) -G^{A}_{\eta, ab}(\mathbf{p},\omega)  ].~~~~
\end{eqnarray}
 
We are now in the position to find the dispersion relation of the valley waves. 
We apply the random phase approximation approach and seek the poles of susceptibility
$\mathrm{det}[ 1 - \lambda \Pi(\mathbf{q}, \omega)] =0$, where the generalized intervalley polarization function
\begin{align}\nonumber\label{PolarizationFunction}
&\Pi_{ab,cd} (\mathbf{q}, \omega) = \frac{ig}{2}\int
\bigg[ 
G^{R}_{-, ab}(\mathbf{p}+\mathbf{q},\Omega+\omega) G^{K}_{+, cd}(\mathbf{p},\Omega)\\
&+ G^{K}_{-, ab}(\mathbf{p}+\mathbf{q},\Omega+\omega) G^{A}_{+, cd}(\mathbf{p},\Omega)
\bigg]\frac{d^2pd\Omega}{(2\pi)^3}
\end{align}
is a $4\times 4$ matrix due to the direct product of two sublattice Pauli matrices. 

Substituting Green functions from Eqs. (\ref{RA}) and (\ref{K}) into Eq. (\ref{PolarizationFunction}) and performing the integration over frequency, one obtains
\begin{align}\label{Pi}\nonumber
&\Pi_{ab,cd} (\mathbf{q}, \omega) = \frac{g}{4}\sum_{s,s'}\int \frac{d^2 p}{(2\pi)^2} \left(\sigma^{0}_{ab} - s \bsigma^*_{ab} \cdot \frac{\mathbf{p}+\mathbf{q}}{| \mathbf{p}+\mathbf{q}|} \right)
\\
&\times \left(\sigma^{0}_{cd} + s' \bsigma_{cd} \cdot \hat{\mathbf{p}} \right) \frac{ f_{-}[E_{s', -}(\mathbf{p}+\mathbf{q})] -f_{+}[E_{s, +}(\mathbf{p})] }{\omega - E_{s',-}(\mathbf{p}+\mathbf{q}) +E_{s,+}(\mathbf{p}) + i\delta}.
\end{align} 
The chemical potential shift $\delta\mu$ enters only via the Fermi distribution functions, while the exchange energy $|\lambda n|$ is the spectral property.

To proceed, one introduces the sublattice pseudospin representation $\tilde{\Pi}^{i,j}(\mathbf{q}, \omega) = \frac{1}{2}\textrm{tr}\sigma^{i}_{bc}\Pi_{ab,cd}(\mathbf{q}, \omega) \sigma^{j}_{da}$, where indices $i,j$ take the values $(0, x, y, z)$ and $\textrm{tr}$ is the trace of Pauli matrices. Evaluating the integral in Eq. (\ref{Pi}) at zero temperature, one obtains
\begin{align}\nonumber\label{New_representation}
&\tilde{\Pi} (\mathbf{q}, \omega) = \frac{v k_0 - \mu}{4\pi v^2}\mathrm{diag}(1,2,0,1)
\\
&- \frac{\mu}{2\pi v^2} \left\langle \frac{\lambda n +\delta\mu + v \mathbf{q}\cdot \hat{\mathbf{p}} }{\omega - \lambda n - v \mathbf{q}\cdot \hat{\mathbf{p}} + i\delta} \mathcal{M}(\hat{\mathbf{p}}) \right\rangle,
\end{align}
where the terms on the first and and second lines describe the inter and intra-band contributions, respectively,  the momentum cutoff $k_0$ is of the order of the inverse lattice spacing, which is introduced so that the integral in Eq. (\ref{Pi}) converges in the ultraviolet, $\mathrm{diag}()$ defines the diagonal matrix, $\langle...\rangle$ defines the integration over the directions of momentum, and
\begin{equation}\label{MMM}
\mathcal{M}(\hat{\mathbf{p}}) = \left(
 \begin{matrix}
  \hat{p}_y^2& 0& \hat{p}_y& -i \hat{p}_x \hat{p}_y\\
 0& 0& 0&0\\
 \hat{p}_y& 0& 1& - i \hat{p}_x\\
 i \hat{p}_x \hat{p}_y& 0& i \hat{p}_x & \hat{p}_x^2
 \end{matrix}
\right).
\end{equation}
Here, small contributions $\propto vq/\mu \ll 1$ are neglected, which results in zero matrix elements $\mathcal{M}_{i,x}=\mathcal{M}_{x,i}$. This means that the relaxation rate of the corresponding collective mode is of the order of $\mu \gg |\omega|$.
Generally, the position of zero elements in matrix Eq. (\ref{MMM}) depends on the sign of the helicity operator in the Dirac Hamiltonian Eq. (\ref{Graphene_Hamiltonian}).
\begin{figure}[t]
\begin{tabular}{c}
\centering
\includegraphics[width=7cm]{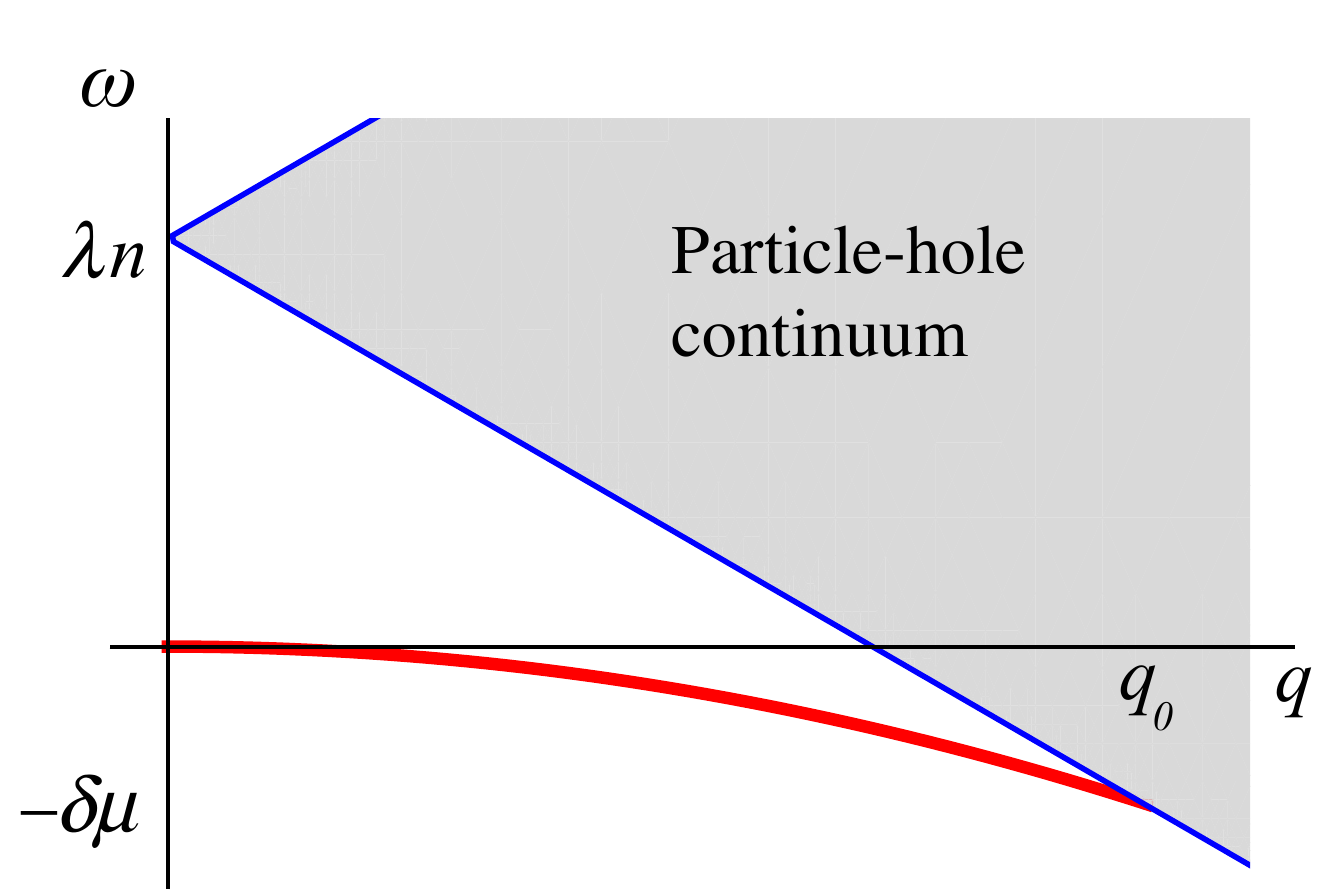}
 \end{tabular}
\caption{Dispersion relation $\omega(q)$ of the collective mode (shown in red) in the case of a positive interaction constant $\lambda>0$. Dispersion crosses the boundary of the electron-hole continuum (shown by the light-gray region) at frequency $\omega_0 = - \delta\mu$ and wave vector $q_0 = 2\pi v|n| /\mu$.}
\label{fig2} 
\end{figure}

Assuming $|\omega| < |\lambda n|$, $vq< |\lambda n|$, $\mu \ll vk_0$, and neglecting excitonic instabilities, we find two modes with a gap
$\omega = \frac{\lambda n}{2} (1- \frac{\lambda (vk_0-\mu)}{2\pi v^2}  )/(1- \frac{\lambda (vk_0-\mu)}{\pi v^2} )$ at $q=0$.
The matrix in Eq. (\ref{New_representation}), which describes the interband contribution, contains a zero diagonal component. It gives rise to 
a gapless mode with the dispersion relation
\begin{equation}\label{Dispersion2D}
\omega = - \left(1- \frac{\lambda \mu}{2\pi v^2} \right) \frac{v^2 q^2}{2 \lambda n}.
\end{equation}
From the denominator of the second term in Eq. (\ref{New_representation}), it follows that
the valley wave becomes damped when the dispersion relation crosses the boundary of the electron-hole continuum, defined by $\lambda n + vq > \omega > \lambda n - vq$, at frequency $\omega_0 = - \delta\mu$ and 
wave vector $q_0 =  2\pi v |n| / \mu$. Dispersion relation of the valley wave is shown in Fig. \ref{fig2}.

\emph{Spin waves in a 2D Dirac semimetal with nonequilibrium spin orientation}.
Let us now briefly comment on the collective modes in a 2D Dirac semimetal with the spin imbalance focusing on graphene as a material candidate.
The electrical spin injection in graphene is well studied experimentally \cite{Spin_injection_first, spintronics_graphene_large, spintronics_review}. 
It was shown that the spin diffusion length is of the order of several micrometers and the spin-relaxation time is up to a few nanoseconds \cite{spintronics_review}. 
The large spin accumulation, of the order of several $\textrm{meV}$, which allows us to study the spin wave modes in this material. 

We consider the case when the spin imbalance in two valleys is the same and focus on the intra-valley spin wave excitations. It suffices to consider
contributions from both valleys separately. In such a case the system is in an energy equilibrium state but with unequal population of spin states. The Hamiltonian is given by
\begin{eqnarray}
H_{\pm, ab}(\mathbf{p}) = v(\sigma^x_{ab} p_x + \sigma^y_{ab} p_y) \mp \frac{\lambda n}{2} \sigma^0_{ab},
\end{eqnarray}
where $\pm$ denotes the two spin states of a single valley. 

We look for a response of the nonequilibrium spin polarization in the system to the transverse magnetic field $\mathcal{H}_{+}(\mathbf{r},t) = (\mathcal{H}_x + i \mathcal{H}_y) e^{i \mathbf{q}\cdot\mathbf{r} - i \omega t}$. The derivation of the poles of the transverse spin susceptibility exactly follows the calculations given in the previous section with a formal substitution of the valley to spin index in the polarization function Eq. (\ref{PolarizationFunction}). Magnetic states due to Stoner instabilities are not considered. Hence, one has to substitute the following in expression (\ref{New_representation}): $\mathrm{diag}(0,1,1,2)$ for the interband contribution and
\begin{equation}
\mathcal{M}(\hat{\mathbf{p}}) = \left(
 \begin{matrix}
  1& \hat{p}_x & \hat{p}_y & 0\\
 \hat{p}_x & \hat{p}_x^2 & \hat{p}_x \hat{p}_y & 0\\
 \hat{p}_y & \hat{p}_x \hat{p}_y & \hat{p}_y^2 & 0\\
 0 & 0 & 0 & 0
 \end{matrix}
\right)
\end{equation}
for the intraband contribution. As a result, one arrives at the dispersion relation of the gapless spin wave mode, for which one has to formally substitute the intervalley interaction constant $\lambda$ and 
particle density difference between two valleys $n$ with corresponding intravalley spin analogs in Eq. (\ref{Dispersion2D}). The dispersion relation of the spin waves is gapless provided the spin-orbit interaction and Zeeman energy of the magnetic field are zero. A similar expression for the dispersion relation of spin waves in metals and semiconductors with nonequilibrium spin orientation was given in Ref. \cite{Aronov}. Here we show that the Dirac semimetals with a population imbalance can host spin waves as well. 

\emph{Collective modes in a 3D semimetal with population imbalance}.
Let us now comment on the collective modes in a 3D Dirac or Weyl semimetal.
The minimal model with four valleys in the absence of the electron-electron interaction can be described by the Hamiltonian
\begin{equation}\label{3D_hamiltonian}
H_{\sigma,\eta}(\mathbf{p}) = \sigma \eta \bsigma \cdot \mathbf{p},
\end{equation}
where $\sigma=\pm1$ and $\eta = \pm 1$ are spin and orbital indices, $\bsigma = (\sigma^{x}, \sigma^{y}, \sigma^{z})$, and momentum is counted with respect to the corresponding Dirac point. The sign of $\sigma\eta$ determines the chirality of the Weyl fermion.

The matrix components in Eq. (\ref{New_representation}) describing the inter-band contribution in the 3D case can be found from $ \textrm{tr}\left(\sigma^{0}_{ab}\sigma^{0}_{cd} - \sigma \eta  \sigma' \eta' \frac{1}{3}\bsigma_{ab} \cdot \bsigma_{cd} \right)\sigma^{i}_{bc} \sigma^{j}_{da}$. The important difference with the above considered 2D semimetal case is that this diagonal matrix has a zero component for $i = j = 0$ provided $\sigma \eta  \sigma' \eta' = 1$.
Hence, in the 3D case the gapless mode exists only if  the coupling is between the states of the same chirality. This leads to an observation that the ferromagnetic Weyl semimetal with only two Weyl cones in the band structure cannot host a gapless intervalley mode.
This is in contrast with the 2D case, where the diagonal matrix in Eq. (\ref{New_representation}) has a zero component for any helicity of the Dirac fermion.

The population imbalance in the 3D Dirac semimetal might be induced, for example, by the effect of the chiral anomaly, which will generate the chiral charge density $n$ proportional to the product of the electric and magnetic fields $\delta\mu \sim \mathbf{E}\cdot \mathbf{B}$.

In Eq. (\ref{3D_hamiltonian}), introducing the electron - electron exchange interaction $H_{\sigma,\eta}(\mathbf{p})  \rightarrow H_{\sigma,\eta}(\mathbf{p}) - \sigma\eta \lambda n/2$, where $\lambda$ is now the Fourier component of the screened potential of the exchange interaction in 3D, the shift of the chemical potential $\mu_{\sigma \eta} = \mu + \sigma \eta \delta\mu/2$, and following the steps described in the previous section, one finds a gapless mode with the dispersion relation at $|\omega|, vq < |\lambda n|$,
\begin{equation}\label{Dispersion3D}
\omega = - \left(1- \frac{\lambda \mu^2}{4\pi^2 v^3} \right) \frac{v^2 q^2}{3 \lambda n}.
\end{equation}
The mass of the valley wave shall be proportional to the product $\mathbf{E}\cdot \mathbf{B}$ in the case of the chiral anomaly. A collinear orientation of the fields supports the condition for the formation of collective waves.  

\emph{Amplification}. 
The excitation of valley and spin waves lowers the energy of the system, which is described by the minus sign in the dispersion relations, Eq. (\ref{Dispersion2D}) and Eq. (\ref{Dispersion3D}), provided $\lambda >0$.
The valley and spin waves generated in the presence of the population imbalance serve as an energy gain source for their external perturbation.
However, in the case $\lambda< 0$, the dispersion relation reads $\omega \propto v^2q^2/|\lambda n|$, and the collective mode serves as an energy bath.
To specify, let us consider the case of spin waves. The dissipation of the magnetic field is described by a time average 
\begin{eqnarray}
Q(\omega) = \frac{1}{4\pi}\langle \mathcal{H} \partial_t \mathcal{B}\rangle \propto | \mathcal{H}|^2 \omega \textrm{Im}\chi(\omega),
\end{eqnarray}
where $\chi \propto - \lambda n [\omega - \omega_0 + i\gamma]^{-1}$ is the transverse magnetic susceptibility and $\gamma>0$ describes the relaxation rate of the collective mode (for example, due to disorder \cite{Zyuzin_Zyuzin}) with the dispersion relation given by $\omega_0(\mathbf{q})$. Hence, at the resonance frequency $\omega = \omega_0$, one has $\mathrm{sgn}[Q(\omega_0)] = \mathrm{sgn}[\lambda n \omega_0] $. For $Q<0$, one expects dips in the absorption spectra in the spin-wave resonance experiment.

To summarize, we calculate the dispersion relation of the inter-valley collective mode in two and three dimensional Dirac and Weyl semimetals in the presence of a valley population imbalance. The 
spectrum is gapless, quadratic in the wave vector, and with a mass determined by the relative valley charge density difference.
In the 3D case, such a mode exists provided the coupling is between a pair of Weyl cones of the same chirality. 
The spin wave analog is discussed for the case of a 2D semimetal with the non-equilibrium spin orientation. Collective modes 
serve as an energy gain source for the external field, that generates transitions between the valley or spin states.

We are grateful for the hospitality of the Pirinem School of Theoretical Physics. 
A.A.Z. acknowledges the support by the Academy of Finland (Project No. 308339).
\bibliography{Imbalance_references}
\end{document}